\documentclass[aps,superscriptaddress,prl,reprint]{revtex4-2}
\label{pre:document-class}

\label{pre:packages}
\usepackage{siunitx}
\usepackage[normalem]{ulem}
\usepackage{graphicx}
\usepackage{amsmath,amssymb}
\usepackage[T1]{fontenc}

\usepackage[inline]{enumitem}
\usepackage{array}
\usepackage[x11names]{xcolor}
\usepackage{comment}

\usepackage{hyperref}

\hypersetup{
  colorlinks=true,
  citecolor=blue,
  linkcolor=blue,
  urlcolor=blue,
}

\usepackage{pifont}
\usepackage[capitalize]{cleveref}
\usepackage[caption=false]{subfig}

\usepackage{multirow}

\usepackage{svn-multi}
\svnidlong
{$LastChangedBy$}
{$LastChangedRevision$}
{$LastChangedDate$}
{$HeadURL$}

\newif\ifsol
\label{pre:ifsol}
\solfalse

\usepackage{tabularx}
\newcolumntype{Y}{>{\centering\arraybackslash}X}

\ifsol

  \usepackage[prefix=]{xcolor-solarized}
  \colorlet{gray}{base01}
  \colorlet{hl}{base02}
  \makeatletter
  \newcommand{\globalcolor}[1]{%
    \color{#1}\global\let\default@color\current@color
  }
  \makeatother
  \AtBeginDocument{
    \globalcolor{base0}
    \pagecolor{base03}
  }

\else

  \usepackage{xcolor}
  \colorlet{hl}{green!50!}
  \colorlet{green}{green!50!black!}
  \colorlet{blue}{blue!55!black!}

\fi

\label{pre:macros}

\renewcommand{\vec}[1]{\mathbf{#1}}

\mathchardef\mhyphen="2D

\newcommand{\tn}{T_{\mathrm{N}}}

\newcommand{\tmi}{T_{\mathrm{MI}}}

\newcommand{\redsout}{\bgroup\markoverwith{\textcolor{red}{\rule[0.4ex]{4pt}{1pt}}}\ULon}

\begin{document}

\label{sec:title}
\title{Intrinsic single crystals of MnTe altermagnet
}

\label{sec:authorlist}

\author{Kim-Khuong Huynh${}^\ddag$}
\email{hkkhuong@chem.au.dk}

\author{Michael Anthony Quintero${}^\ddag$}

\affiliation{Department of Chemistry, Aarhus University, Langelandsgade 140, 8000 Aarhus C, Denmark}

\thanks{These authors contributed equally to this work.}
\author{Martin Klanj\v{s}ek}
\affiliation{Jozef Stefan Institute, Jamova cesta 39, 1000 Ljubljana, Slovenia}

\author{Tilen Knafli\v{c}}
\affiliation{Jozef Stefan Institute, Jamova cesta 39, 1000 Ljubljana, Slovenia}

\affiliation{Silicon Austria Labs, Europastra\ss{}e 12, 9524 Villach, Austria}

\author{Yuta Ishii}
\author{Norimasa Sasabe}
\affiliation{Center for Basic Research on Materials, National Institute for Materials Science (NIMS), Tsukuba 305-0047, Japan}

\author{Nhu-Quynh T. Phan}

\author{Frej S{\o}ren Rattenborg}
\affiliation{Department of Chemistry, Aarhus University, Langelandsgade 140, 8000 Aarhus C, Denmark}


\author{Yuichi Yamasaki}
\affiliation{Center for Basic Research on Materials, National Institute for Materials Science (NIMS), Tsukuba 305-0047, Japan}
\affiliation{International Center for Synchrotron Radiation Innovation Smart, Tohoku University, Sendai 980-8577, Japan}
\affiliation{Center for Emergent Matter Science (CEMS), RIKEN, Wako 351-0198, Japan}

\author{Denis Ar\v{c}on}
\affiliation{Jozef Stefan Institute, Jamova cesta 39, 1000 Ljubljana, Slovenia}
\affiliation{Faculty of Mathematics and Physics, University of Ljubljana, Jadranska cesta 19, 1000 Ljubljana, Slovenia}

\author{Bo Brummerstedt Iversen}
\email{bo@chem.au.dk}
\affiliation{Department of Chemistry, Aarhus University, Langelandsgade 140, 8000 Aarhus C, Denmark}

\date{\today}

\begin{abstract}
  We report the synthesis methodology, structure, and intrinsic properties of ultra-high quality single crystals of MnTe, an archetypal altermagnet.
  The crystals, obtained from self-flux method, are nearly free from crystal imperfections and disproportionate chemical compositions as seen by various investigation methods.
  In measurements under quasi free-standing configuration minimizing stress induced effects, the crystals exhibit complex and anisotropic domain kinetics in both superheating and supercooling regimes around the altermagnetic transition at $\tn = \SI{310}{\kelvin}$.
  An Anderson insulating state is observed below $T_{\mathrm{MI}}\approx \SI{150}{\kelvin}$ with a carrier density of about $\SI{1.6e17}{\per \cubic \centi\meter}$, being sharply contrast to metallic states usually seen in Te-deficit samples.  
  Nevertheless, hallmarks of altermagnetism, anomalous Hall effect and X-ray magnetic circular dichroism signal, are robust in this intrinsic limit, however with significantly reduced magnitudes.
\end{abstract}

\maketitle

\paragraph{Introduction --}
Altermagnetism (ALTM), a newly found antiferromagnetic-like order which breaks time-reversal symmetry with strictly zero net magnetization, has quickly entered the spotlight of condensed matter physics and materials science \cite{smejkal2022,smejkal2022a}.
The clarity of symmetry analyses pushes forward a deeper recognition of existing, well-documented antiferromagnetic (AFM) orders \cite{smejkal2022,smejkal2022a,cheong2025} and naturally opens further discussions on the intricate microscopic mechanisms \cite{mcclarty2024,sasabe2025,mazin2024}, the magnitudes and investigating methodologies \cite{sasabe2025,mazin2024,takahashi2025}, and material designs \cite{fender2025}.
The achieved fundamental understanding and potential of applications motivate great efforts in the field.

\begin{figure}
  \centering
  \includegraphics[width = \linewidth]{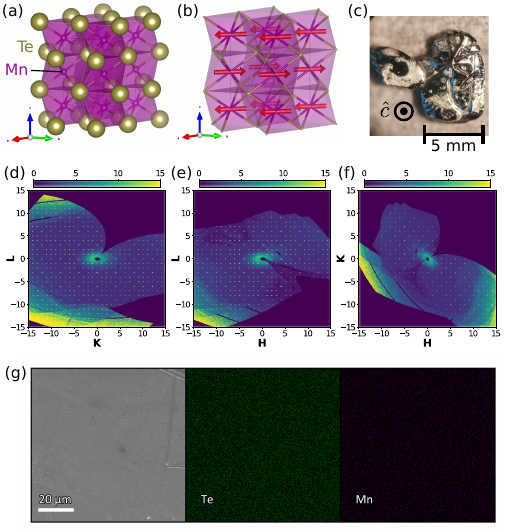}
  \caption{
    Crystal quality.
    (a) Crystal structure of MnTe as obtained from the refinement of synchrotron SCXRD data.
    (b) AFM order with spin axis aligned along $[010]$-direction.
    (c) Single crystals of MnTe obtained from flux growth, after washing by aqua regia.
    (d-f) Reconstructed SCXRD precession images oriented along the $[10\bar{1}0]$, $[01\bar{1}0]$, and $[0001]$ directions.
    (g) SEM-EDX result showing homogeneous distribution of stoichiometric Mn:Te composition.
  }
  \label{fig:crystallinity}
\end{figure}

Among many candidates, MnTe [Fig.~\ref{fig:crystallinity}(a-b)] has emerged as an archetypal ALTM \cite{mazin2024,mazin2023}, in which key spectroscopic and anomalous transport signatures of the spin-dependent electronic structures have been experimentally established \cite{gonzalezbetancourt2023,krempasky2024,osumi2024,hariki2024,amin2024,takegami2025}.
These observations, while constituting a proof-of-concept for ALTM, immediately raise critical questions.
Most notably , MnTe is an AFM charge-transfer insulator \cite{zaanen1985,khomskii2014,mazin2024} with a large band gap suppressing static electrical conductivity in the intrinsic limit.
At the same time, its chemistry renders it highly susceptible to Te-deficiency, which introduces holelike carriers via self-doping \cite{wasscher1969,svane2025}.
Such non-stoichiometry, commonly encountered in vapor-transport-grown samples \cite{ wasscher1969,kriegner2017,krempasky2024,gonzalezbetancourt2023,bey2024,kluczyk2024}, can pin the Fermi level ($E_{\mathrm{F}}$) near regions of large ALTM Berry curvature located well below the valence band maximum \cite{gonzalezbetancourt2023}.
While this effect can facilitate the observation of the anomalous Hall effect (AHE) via self-doping, Te-vacancies also introduce inhomogeneous perturbations to the local environment of the magnetic Mn$^{2+}$ ions, the consequences of which are still unknown.
Furthermore, because $E_{\mathrm{F}}$ is preselected by the defects, tuning it via electrostatic gating is difficult, hindering the study of energy dependence of ALTM properties.
The  chemical/structural uncertainties and the absence of ALTM hallmarks in clean, well-investigated MnTe crystals closely parallel earlier challenges in high-$T_{\mathrm{C}}$ cuprate superconductors, superconducting phases in doped topological insulators, and the quantum spin liquid candidate RuCl$_3$, where hidden material variables obscured intrinsic behaviors \cite{armitage2010a,kevy2021,ojeda-aristizabal2025}.

Mechanical stress is equally impactful in MnTe due to its substantial piezomagnetic effect \cite{aoyama2024}.
Whereas stress can enhance ALTM signals by reducing the population of domains hosting opposite order parameters, the observed magnitudes always depend on the specific domain and stress conditions  \cite{amin2024,takegami2025}. 
It follows that in order to observe the intrinsic properties of MnTe, crystals should be not only chemically precise, but also free-standing mechanically.
The large AHE initially observed in thin film samples \cite{wasscher1969,kriegner2017,krempasky2024,gonzalezbetancourt2023} was later found to be strongly dependent on the substrate material \cite{bey2024}, a direct consequence of the AHEs sensitivity to epitaxial strain.
We are thus motivated in studying high-quality single crystals of MnTe which satisfy the chemical, structural, and mechanical criteria for intrinsic behaviors.

In this Letter, we report the synthesis and the structure, and properties of such defect- and stress-free MnTe single crystals.
In quasi free-standing configurations, we observed complex effects arising from short-ranged magnetic order and evolution of domain structures, which manifest in terms of complex anisotropic hysteresis loops in both superheating and supercooling regimes around the ALTM transition occurring at $\tn\approx \SI{310}{\kelvin}$.
These metastabilities suggest an intricate between magnetic and mechanical energy in MnTe.
The ALTM domains can be partially controlled by magnetic fields perpendicular to the $c$-axis, as evident by our nuclear magnetic resonance (NMR) measurements.
An Anderson insulating state is observed at low $T$, being consistent with the low carrier density associated with stoichiometric composition of the crystals.
Both anomalous Hall effect (AHE) and X-ray magnetic dichroism (XMCD) signals, hallmarks of ALTM, are present in this clean limit, however with greatly reduced magnitudes in comparison with those observed in thin films.
ALTM properties are not only robust but also sensitive to the carrier density and mechanical boundary conditions, making the obtained MnTe crystals an ideal platform for electrostatic and mechanical tuning ALTM.

\paragraph{Crystal quality --}
\label{sec:crystal-quality}


We employed a tailored Te-based flux synthesis to grow single crystals of MnTe \footnote{See End Matter}.
The obtained crystals are well-formed, thick platelets, with lateral dimensions in the $ab$ plane typically over about $\SI{8}{\milli\meter}$ and a thickness along $c$-axis reaching $\SI{1.5}{\milli\meter}$ [Fig.~\ref{fig:crystallinity}(c)].
Synchrotron single crystal X-ray diffraction (SCXRD) measurements produced data with an excellent internal consistency index $R_{\mathrm{int}} \approx \SI{4.38}{\percent}$.
In the reconstructed precession images, Fig.~\ref{fig:crystallinity}(d-f), all reflections appear as bright, round spots and are well-indexed by the space group $P6_3/mmc$.
Importantly, there is no detectable diffuse scattering or mosaicity, indicating that the measured crystals are free from both strains and impurity phases.
The diffraction data are exceptionally well-described by the structural model shown in Fig.~\ref{fig:crystallinity}(a), with quality index of the model fit $R_1 \approx \SI{0.98}{\percent}$.
No signature of structural transition was observed within the temperature ($T$) window of $\SI{40}{\kelvin} \leq T \leq \SI{300}{\kelvin}$.

\begin{figure}
  \centering
  \includegraphics[width = .45\textwidth]{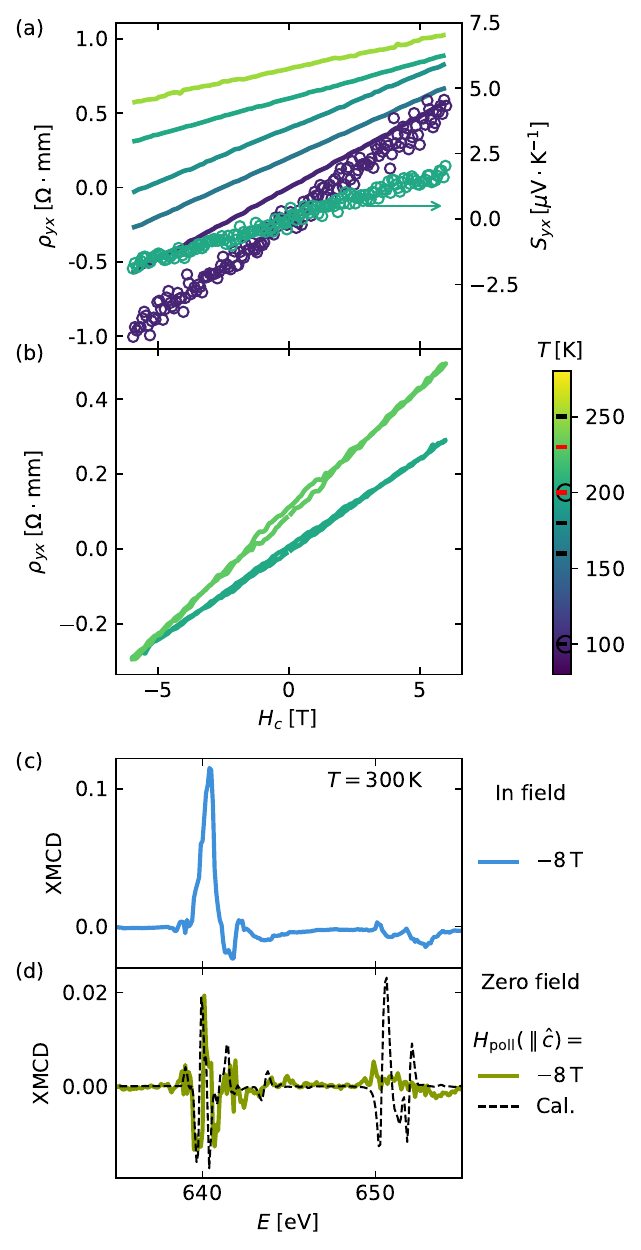}
  \caption{
    ALTM hallmarks.
    (a) Hall ($\rho_{yx}$) and Nernst ($S_{yx}$) effects at various temperatures.
    (b) Hysteresis loops associated with AHE observed in $\rho_{ya}(H)$ curves measured at $T = \SI{200}{\kelvin}$ and $\SI{230}{\kelvin}$.
    A line (circle) on the color bar on the right indicates the temperature at which a $\rho_{yx}$ ($S_{yx}$) curve was measured.
    (c) Room temperature XMCD signals collected in $\vec{H}\parallel c$ exhibit clear $\vec{H}$-induced magnetizations.  
    (d) Zero-field XMCD signals after polling at $H_{\mathrm{poll}} = \SI{-8}{\tesla}$ showing the ALTM spectral shape predicted by theory.
  }
  \label{fig:absence}
\end{figure}

Elemental composition analyses using scanning electron microscopy and energy dispersive X-ray spectroscopy (EDS) [Fig.~\ref{fig:crystallinity}(g)] confirm the stoichiometric ratio of MnTe and a uniform distribution over the crystal \cite{Note1}.
The composition of the current crystals is free from Te-deficiency, being sharply contrast the MnTe$_{1-\delta}$, with $\delta \approx 0.05$ common to vapor-grown samples.
Hall effect measurements [Fig.~\ref{fig:absence}(a)] show that the carrier density in our crystals is about $\SI{1.6e17}{\per\cubic\centi\metre}$, being more than one order of magnitude smaller than the common value in other samples \cite{gonzalezbetancourt2023,kluczyk2024}.
The small carrier number is consistent with the stoichiometric composition and the insulating behavior shown below.
Finally, the experimental X-ray absorption spectrum \cite{Note2} is consistent with theoretical calculations of a dipole transition $2p^{5}3d^{6} \rightarrow 2p^{5}3d^{6}$ of Mn$^{2+}$ ions in $D_{3d}$ symmetry \cite{Note1}.

\paragraph{Complex ALTM/AFM transition --}
\label{sec:phys-prop}

\begin{figure*}
  \centering
  \includegraphics[width = \textwidth]{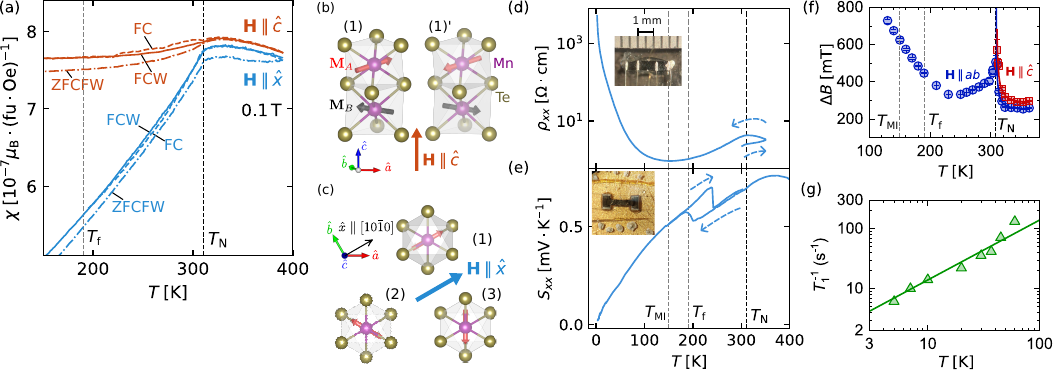}
  \caption{
    Complex ALTM/AFM transition, domain degeneracy, and low-$T$ insulating state.
    (a) Anisotropic superheating and supercooling hysteresis loops around $\tn$ exhibited by magnetic susceptibilities $\chi_{cc}(T)$ and $\chi_{xx}(T)$.
    The magnetic field parallel to $c$- and $x$-axes, as shown in (b) and (c), respectively.
    (b) and (c) illustrate the $c$- and in-plane degeneracy of domain orientations, respectively.
    The spin canting angle in (b) is exaggerated for clarity.
    (d-e) Hysteresis loops around $\tn$ and Anderson localization at low-$T$ in the zero-field $T$-dependencies of resistivity $\rho_{xx}(T)$ and the Seebeck coefficient $S_{xx}(T)$.
    The dashed arrows indicate the cooling/warming processes, and the insets show the set-up of the measurements.
    (f) Temperature dependence of the EPR linewidth for magnetic field in the $ab$ plane (blue circles) and perpendicular to the $ab$ plane (red circles).
    The solid blue and red lines are fits to high-temperature critical broadening, when approaching the N\'eel transition from above.
    (g) Temperature dependence of the $^{55}$Mn spin-lattice relaxation rate $T_1^{-1}(T)$, which follows a Korringa-like linear behaviour (indicated by the line).
  }
  \label{fig:complex}
\end{figure*}

Fig.~\ref{fig:complex}(a) displays the magnetic susceptibilities $\chi_{cc}(T)$ and $\chi_{xx}(T)$ measured in magnetic fields $\vec{H}_c$ and $\vec{H}_x$ parallel to $c$- and $x$-axes, respectively.
Here $x$-axis is parallel to the $[10\bar{1}0]$-direction of the crystal [Fig.~\ref{fig:complex}(c)].
Both susceptibilities exhibit complex metastable behaviors in both superheating ($T > \tn \approx \SI{310}{\kelvin}$) and supercooling ($T < \tn$) regimes.
The thermal cycles in Fig.~\ref{fig:complex}(a) started by zero-field-cooling the sample from $T = \SI{370}{\kelvin} > \tn$ to $\SI{100}{\kelvin}$, at which $H=\SI{0.1}{\tesla}$ was applied, and the susceptibilities were measured during field-warming (ZFCFW), the field-cooling (FC), and eventually field-warming after field-cooled (FCW).
ZFCFW and FC susceptibilities bifurcate at temperatures high in the superheating regime, and FC and FCW hysteresis loops also manifest in the supercooling regime.

ZFCFW-FC bifurcations are driven by the selection of the most energetically favorable configuration of magnetic domains and usually occur at the vicinity of the transition.
In MnTe, because the bifurcations emerge at $T>\tn$ [Fig.~\ref{fig:complex}(a)], such optimal ALTM/AFM domain structure is likely to be pre-selected early in the superheating regime.
This effect is also prominent at zero-field as shown by the $\rho_{xx}(T)$ curve in Fig.~\ref{fig:complex}(d),
in which the initial ``low-$T$ state'' at $T = \SI{300}{\kelvin}$ can be brought into a more resistive ``high-$T$ state'' by warming it to $\SI{350}{\kelvin}$.
In understanding these effects, it is important to note that the superheating regime of MnTe is not paramagnetic, but is dominated by short-range magnetic correlations with the length scale of about $\SI{20}{\angstrom}$, as shown by a recent neutron study \cite{baral2022}.
Heat treatment and application of magnetic field on the short-range correlations appear to have lasting memory effects on the subsequent evolutions of the domain states \cite{Note1} .

Within the supercooling regime, the presence of the hysteresis loops between the FC-FCW curves [Fig.~\ref{fig:complex}(a)] and in the zero-field Seebeck coefficient $S_{xx}(T)$ indicate that domain kinetics continues to be active.
Similar signatures also appear in the $T$-dependencies of the in-plane EPR linewidth $\Delta B$ [Fig.~\ref{fig:complex}(f)] and asymmetric parameter $\alpha$, as well as in the specific heat \footnote{See Supplementary Materials}.
As shown below, the effects of magnetic fields on domains are anisotropic.
On the other hand, the transport properties shown in Fig.~\ref{fig:complex}(d-e) are sensitive to the overall evolution of the domain structures.
The low-$T$ closing of the $S_{xx}$-hysteresis at $T_{\mathrm{f}} \approx \SI{190}{\kelvin}$ can be defined as the freezing temperature, at which the domain structure ceases or starts to evolve upon further cooling or warming.

\paragraph{Anisotropic domain effects --}
Below $\tn$, the volume of a MnTe crystal can be fragmented into six possible types of domains as dictated by the sixfold degeneracy of the N\'eel vector, $\vec{L} \equiv \vec{M}_A -\vec{M}_B$.
As shown by Fig.~\ref{fig:complex}(b-c), two orders of degeneracy come from the time-reversal transformation, $(i) \leftrightarrow (i)^{\prime}$, and three from the in-plane $2\pi/3$-rotations, $(i), (i)^\prime \leftrightarrow (j), (j)^\prime$, with $i, j = 1, 2, 3$ \cite{komatsubara1963,kriegner2017,aoyama2024}.
The sublattice magnetizations of $(i)$ and $(i)^\prime$ cant along opposite directions of $c$-axis \cite{aoyama2024}, and $\vec{H}_c$ favors domains  $(i)$ having parallel canting over their time-reversal copies [Fig.~\ref{fig:complex}(b)].
Therefore, cooling in $\vec{H}_c$ selects $(i)$'s as the major species of domains.
Because volume fractions of $(i)$ and $(i)^\prime$ are more balanced in the zero-field-cooled state, this domain selection leads to $\chi_{cc}^{\mathrm{FC}} > \chi_{cc}^{\mathrm{ZFCFW}}$.
On the other hand, a larger fractional difference between $(i)$ and $(i)^\prime$ corresponds to a higher mechanical stress exerted on the crystal because of piezomagnetic effect.
The balance between magnetic and stress energies might be the mechanism for $\chi_{cc}^{\mathrm{FC}}$ to decrease and merge with $\chi_{cc}^{\mathrm{FCW}}$ as $T \rightarrow T_{\mathrm{f}}$.

Whereas $\vec{H}_c$ does not directly affect in-plane domain structure, $\vec{H}_x$ splits the in-plane threefold degeneracy and leaves the time-reversal doublets untouched \cite{komatsubara1963}.
Domains of types $(1)$ and $(1)^\prime$ have their common in-plane spin-axis parallel to $\vec{H}_x$, and their magnetic energy is larger.
The other four types have their original spin-axes oriented $\pm\SI{120}{\degree}$ away from $\vec{H}_x$, and thereby are more favorable energetically [Fig.~\ref{fig:complex}(c)].
The angles also make the spins more susceptible to the alignment effect of $\vec{H}_x$, leading to a larger $\chi_{xx}$ \cite{yosida1996}.
A $\vec{H}_x$-cooled crystal is thus dominated by the domains of $(2)$ and $(3)$ families, leading to $\chi_{xx}^{\mathrm{FC}} > \chi_{xx}^{\mathrm{ZFCFW}}$.
The narrow FC-FCW hysteresis loop in $\chi_{xx}$ suggests a small activation energy for in-plane domain restructuring.

\begin{figure}
  \centering
  \includegraphics[width = .5\textwidth]{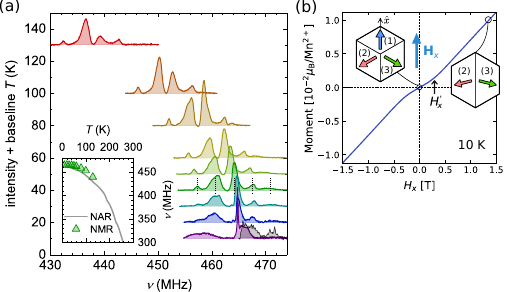}
  \caption{Local probe of magnetic domain.
    (a) The colored bands in the main figure represent the $^{55}$Mn NMR spectra measured in a magnetic field of $\SI{0.5}{\tesla}$ applied in the crystal $ab$ plane at various temperatures.
    The vertical offset of each spectrum corresponds to its measurement temperature.
    For comparison, the spectrum recorded at $\SI{10}{\kelvin}$ and $\SI{0}{\tesla}$ is shown in gray with an enhanced intensity.
    The vertical dashed lines mark the frequencies of the five transitions in the quadrupole-split $I=5/2$ spectrum at $\SI{40}{\kelvin}$.
    The inset compares the temperature dependence of the central-line frequency with the previously reported NAR \cite{walther1971}.
    (b) Magnetization versus $H_x$.
      The cartoons illustrate the depopulating effect of $H_x$ on the domains.
      The arrows represent the directions of the N\'eel vectors in $(1)$, $(2)$, and $(3)$ domains. The time-reversal copies of the shown domain types are omitted for clarity.
  }
  \label{fig:magnetic-resonance}
\end{figure}

\paragraph{In-plane domain depopulation by $\vec{H}_x$--}
\label{sec:domain}
$^{55}$Mn NMR provides a microscopic probe of the local magnetic environment in MnTe and enables direct insight into ALTM domain structure.
As shown in Fig.~\ref{fig:magnetic-resonance}(a), the $^{55}$Mn NMR spectrum initially recorded at zero magnetic field and $T = \SI{10}{\kelvin}$ (plotted in gray) is unexpectedly weak with an irregular shape, a result highly different from the expected quadrupole-split pattern in case of $I=5/2$ nuclear spin \cite{jansa2021}.
Only upon the application of a small in-plane field $H_{ab}^\prime = \SI{0.5}{\tesla}$, the spectrum changes into a well-resolved quadrupole-split pattern of the $^{55}$Mn nucleus at the hexagonal Mn site.
The frequency of the central transition decreases smoothly with increasing temperature and closely follows the nuclear acoustic resonance (NAR) line reported for MnTe \cite{walther1971}, demonstrating the gradual reduction of the internal hyperfine field $H_{\mathrm{hf}}$ from its maximum values of about $\SI{44}{\tesla}$ at $\SI{10}{\kelvin}$.
Because $H_{ab}^\prime \ll H_{\mathrm{hf}}$, the resonance condition is overwhelmingly dominated by $H_{\mathrm{hf}}$.
However $H_{ab}^\prime$  clearly enables the NMR observation.


The drastic change of the NMR spectral shape induced by $H_{ab}^\prime$ is consistent with the slope bend around $H_{x}^\prime \approx \SI{0.3}{\tesla}$ in magnetization curve shown in Fig.~\ref{fig:magnetic-resonance}(b).
Whereas spin-flop would be a possible mechanism, such a transition should cause an abrupt magnetization jump in $\vec{H}_x$ \cite{komatsubara1963,yosida1996} and a significant shift of the frequency of $^{55}$Mn NMR \cite{Note2}, being qualitatively different from the observed gradual change.
On the other hand, the significant role of $H_{ab}$  the lifting in-plane domain degeneracy can not be ignored \cite{komatsubara1963}.

In a stress-free crystal at zero-field, the six degenerate domain types should be equally populated.
The orientation of $\vec{H}_{\mathrm{hf}}$, which defines the nuclear quantization axis, varies between domains.
The nuclear resonant condition is therefore strongly domain-dependent, and the superposition of signals from multiple domains leads to the irregular linewidth and the reduced intensity observed at $H_{ab} = 0$ \cite{Note2}.
Turning on ${H}_{x}$ raises the energy of $(1)$ and $(1)^\prime$ domains, causing them to be consumed by the other four domain types until they are exhausted at around  $H_x^\prime$.
Further increasing the magnetic field will continuously rotate the spin-axes of the surviving domains from $\pm\SI{120}{\degree}$ to perpendicular to $x$ \cite{yosida1996}.
This lifting of degeneracy partially eliminates the superposition and improves the NMR spectrum, while a considerable linewidth of $\SI{2}{\mega\hertz}$ in Fig.~\ref{fig:magnetic-resonance}(a) still reflects a broad distribution of effective local fields originating from the surviving domain types \cite{Note2}.
The role of $\vec{H}_x$ in consolidating ALTM domains agrees well with an X-ray magnetic dichroism imaging study \cite{amin2024}, which show that both $\vec{H}_x$ and micro-patterning are necessary to bring the micro-size sample to a single domain state.

\paragraph{Low-$T$ localization --}
Whereas sintered MnTe tends to be insulating even at room temperature, thin films and crystals of MnTe are bad metals or shows very weak insulating behaviors \cite{kriegner2016,kriegner2016,dzian2025,aoyama2024,krempasky2024,kluczyk2024}.
In a sharp contrast to the existing data, the $\rho(T)$ curve of the present MnTe exhibits a clear metal-insulator transition at $\tmi \approx \SI{150}{\kelvin}$.
At $T < \tmi$, $\rho(T)$ initially follows an activation law until it crosses over into a variable-range-hoping (VRH) regime at $T \lesssim \SI{16}{\kelvin}$ \cite{Note2}.
The VRH behavior of $\rho(T)$ and the monotonic decrease of $S(T)$ [Fig.~\ref{fig:complex}(c)] are consistent with the picture of Anderson localization \cite{mott2012}.
Although the electronic band carrying non-trivial Berry curvature intersects the Fermi level $E_{\mathrm{F}}$ (see next section), MnTe remains insulating due to disorder-induced localization.
No crystallographic disorder is detectable by SCXRD, suggesting that the localization is magnetically driven.

NMR and EPR measurements shed more light into the intrinsic magnetic dynamics and insulating state at low temperatures.
Because the $T$-dependence of the residual X-band EPR linewidth in Fig.~\ref{fig:complex}(f) closely traces that of $\rho(T)$, the dominant spectral weight likely comes from the spin resonance of conducting electron (CESR) residing at $E_{\mathrm{F}}$.
Modeling the signal by asymmetric Dysonian function \cite{dyson1955} yields reasonable change of the asymmetric parameter in the metallic regime $\tmi < T < \tn$.
However, the CESR lineshape becomes anomalous at $T \approx \tmi$, where the electronic states $E_{\mathrm{F}}$ starts to lose their spatial coherence \cite{Note2}.

In the VRH regime, the spin-lattice relaxation rate $T_1^{-1}(T)$ obtained from ${}^{55}$Mn NMR at $T<\SI{60}{\kelvin}$ follows a Korringa-like linear $T$-dependence, with $T_1$ being of the order of $\SI{100}{\milli\second}$ [Fig.~\ref{fig:complex}(g)].
The Korringa behavior implies that the nuclear spins are relaxed via gapless, weakly itinerant low-energy spin excitations, whereas the material is deep in the insulating regime.
This contrast can be naturally explained by Anderson localized states, which are spatially incoherent but can still contribute sufficient spectral weight at $E_{\mathrm{F}}$ to mediate the nuclear relaxations.

\paragraph{Anomalous Hall effect and XMCD --}
\label{sec:absent_AHNE}

Both Hall ($\rho_{yx}$) and Nernst ($S_{yx}$) effects collected in quasi-free settings \cite{Note1} and $\vec{H}_c$ are predominantly linear, as displayed in Fig.~\ref{fig:absence}(a).
On the other hand, Fig.~\ref{fig:absence}(b) reveals weak AHE-related hysteresis loops emerging after the crystals had been either cooled down or warmed up to the measuring temperature under $\vec{H}_{c}$.
Incidental domain selection due to thermal stress under field-cooling may enable the detectable AHE \cite{aoyama2024}.
Unfortunately, signatures of anomalous Nernst effect are undetectable, probably due to the small signal-noise-ratio in the measurement.
A rough estimation of the anomalous Hall conductivity yields $\sigma_{yx}^{\mathrm{AHE}} \approx \SI{3}{\milli\siemens\per\meter}$, i.e., about three orders smaller than the observed value in thin films.

XMCD measurements carried out at $T = \SI{300}{\kelvin}$ yield results consistent with the transport properties.
Recently, several studies have reported finite XMCD signals with spectral features uniquely characteristic of altermagnetic states in compensated antiferromagnets, thereby distinguishing them from conventional ferromagnetic spectra \cite{Sasabe2023,hariki2024,sasabe2025,ishii2026}.
In Fig.~\ref{fig:absence}(c), the in-field XMCD spectra indicate the presence of a weak ferromagnetic-like magnetization along $c$-axis, the origin of which does not directly relate to ALTM \cite{mazin2024}.
In sharp contrast, as shown in Fig.~\ref{fig:absence}(d), the XMCD signals obtained at zero magnetic field (after polling at $H_{\mathrm{poll}} = - \SI{8}{\tesla}$), although having rather low intensity, exhibit the ALTM characteristic spectral shapes predicted by theory.
Interestingly, theory also suggests that the XMCD responses of ALTM in MnTe are robust even at elevated temperature near $\tn$, being consistent to the spectra shown in Fig.~\ref{fig:absence}(d).
This spectroscopic observation is consistent with being consistent with the AHE and demonstrates that room temperature ALTM persists in the present intrinsic MnTe.

\paragraph{Summary--}
\label{sec:summary}
In summary, our study provides a well-controlled synthesis method for MnTe and an understanding of its intrinsic properties, which, to the best of our knowledge, have not been investigated.
The availability of ultra-clean MnTe and experimental observations of intrinsic ALTM hallmarks near room temperature are fundamental for future studies of ALTM via tuning external parameters.
For example, clean MnTe crystal, free from self-doping effects, is an excellent platform for investigating energetic dependence of ALTM Berry curvature via tuning $E_{\mathrm{F}}$ with electrostatic gating.
Our study also emphasizes the significance of complex domain physics in MnTe.
The observed superheating-supercooling temperature window should be related to the energy scale of domain wall motions, which can play the key role in future domain engineering.
Finally, the $^{55}$Mn NMR spectral shape of ALTM-ordered Mn$^{2+}$ moments calls for better theoretical and experimental understanding on how the magnetic ion interacts with its ALTM local environment.
Such knowledge can be put in parallel with other known spectroscopic signatures of ALTM \cite{sasabe2025}.

\section*{Acknowdegement}
\label{sec:acknowdegement}
  The authors are grateful to Caroline B. Eriksen for help in synthesis, L. Jana and T. Nakamura, and to S. Chowdhury and M. Hoesch for stimulating discussions and excellent measurement supports, to T. Kida, Y. Narumi, M. Hagiwara, and to N. Ubrig for helpful discussion and comments. 
  Y.P. Chen is thanked for letting us use his electronic equipments.
  SPring-8 and NanoTerasu synchrotron facilities are thanked for their hospitality and supports.
  The work was supported by the Villum Foundation (25861), the Danish National Research  Foundation (DNRF189), the Carlsberg Foundation (CF24-1527), the Novo Nordisk  Foundation (NNF23OC0085064), the Pioneer Center for Accelerating P2X Materials Discovery (CAPeX, DNRF P3), and the Independent Research Fund Denmark.
  D.A. acknowledges the financial support of the Slovenian Research Agency through P1-0125 grant.



\section*{Author contributions}
\label{sec:author-contribution}
  K-.K.H. conceptualized the study, measured and interpreted the magnetic, thermodynamic, and transport properties, and wrote the original manuscript.
  M.A.Q. designed and optimized the crystal growth and performed the SCXRD measurements and analyses.
  M.K., T.K., and D.A. measured and interpreted the NMR and EPR.
  Y.I., Y.Y., M.A.Q. and K.-K.H measured XAS and XMCD, which were analyzed and interpreted by Y.I., Y.Y., and N.S..
  N.-Q.T.P. measured and analyzed the SEM-EDX data.
  F.S.R. and M.A.Q. measured the magnetic and transport properties.  
  N.S. and Y.Y. carried out the the XAS and XMCD theoretical calculation.
  B.B.I. acquired funding, initialized and supervised the project.
  All the authors performed the formal analyses, wrote, edited, and reviewed the paper.

\bibliographystyle{apsrev4-1}


\bibliography{./04a_MnTe.bib}

\clearpage

\section*{End Matter}
\label{sec:end-matter}

\paragraph{Crystal growth -- }
\label{sec:growth}
The synthesis were carried out at Department of Chemistry, Aarhus University (AU).
All elements were used as obtained from Chempur:  Mn $325$ mesh powder ($\SI{99.9}{\percent}$) and Te pieces ($\SI{99.999}{\percent}$, low oxygen).
The starting reagents were stored in an Ar glovebox with $ {} < 0.1$ ppm H$_2$O and $ < 0.1$ ppm O$_2$.
A mixture with the Mn:Te molar ratio of $6.9:12.7$ were massed in an Ar glovebox and ground together in an agate mortar and pestle until homogeneous.
The ground powder was loaded into a graphite Canfield-crucible and loaded into a fused silica tube cushioned above and below with silica wool. 
The tube opening was wrapped in parafilm, removed from the glovebox, necked down, evacuated to $2\times{10}^{\mathrm{-4}}$ mbar, and flame sealed.
The air exposure time  were less than $5$ minutes during the whole operation.
The sealed tube was wrapped in Kaolwool to improve the thermal stability, inserted into a larger fused silica tube, and placed upright in a Nabertherm box furnace. 
The tube was heated to 940°C at 80°C/h, soaked at 940°C for 24 hours, and slow cooled to 840°C at 0.3°C/h. 
The setup was then removed from the furnace, inverted, and centrifuged at 1500 RPM for 1 min, then allowed to cool to room temperature before opening.
The starting composition and the heating program of the synthesis were designed based on a Mn-Te binary phase diagram \cite{villars}.
The flux-grown crystals were washed by diluted aqua regia to remove the excess flux on the surface before all measurements \cite{Note2}.

\paragraph{Structure and composition studies -- }
\label{sec:struct-comp-invest}
For SCXRD measurements, crystals with sizes of about $\SI{40}\times\SI{40}\times\SI{40}{\cubic \micro \meter}$ were cut from flux-grown crystals while submerged under liquid nitrogen, then glued to glass fibers.
The data were collected at the BL02B1 beamline of the SPring-8 synchrotron, Japan, with an X-ray photon energy of $\SI{50}{\kilo\electronvolt}$ ($\lambda= \SI{0.2490}{\angstrom})$ at $\SI{40}{\kelvin}$ using a Huber four-circle (quarter-$\chi$) goniometer equipped with a Pilatus3 X 1MCdTe (P3) detector. 
The images were converted to the Bruker “.sfrm” format \cite{lennardkrause} and integrated using SAINT-Plus \cite{zotero-item-23362}.
The integrated data were scaled and corrected using SADABS \cite{krause2015}.
Additional datasets at $\SI{100}{\kelvin}$, $\SI{200}{\kelvin}$, and $\SI{300}{\kelvin}$ were collected on a Rigaku XtaLAB Synergy-S diffractometer with an Ag microfocus source ($\lambda= \SI{0.56}{\angstrom}$) outfitted with a HyPix-Arc $\SI{100}{\degree}$ detector. 
The integration, scaling, and absorption corrections of these in-house datasets were carried out using CrysalisPro \cite{Crysalis}. 
All structures were solved using ShelXT \cite{sheldrick2015} and refined within the Independent Atom Model (IAM) in ShelXL \cite{sheldrick2015a} using the Olex2 GUI \cite{dolomanov2009}.

The surface morphology and element composition of MnTe single crystal were analyzed using a TESCAN CLARA field-emission scanning electron microscope equipped with an E-T detector and coupled with an energy dispersive X-ray spectrometer (SEM-EDS) installed at AU.
Prior to the measurements, the single crystal was mechanically polished and cleaned sequentially with acetone and ethanol.
The SEM images were acquired at an accelerating voltage of $\SI{15}{\kilo\electronvolt}$ and a probe current of $\SI{300}{\pico\ampere}$.
EDS measurements were carried out at an accelerating voltage of $\SI{15}{\kilo\electronvolt}$ with a beam current of $\SI{3}{\nano\ampere}$.

\paragraph{Crystal alignment --}
\label{sec:alignment}
For the measurements of anisotropic physical properties, the crystals were oriented by the Synergy-S diffractometer before cutting to transport samples or mounting to the holder of the magnetometer.
Determination of the crystal orientation matrix was done using a Rigaku XtaLAB Synergy-S diffractometer within CrysalisPro.
The crystal was mounted on a MiTiGen loop using grease, and a brief SCXRD data set was collected.
The unit cell was then indexed from the collected frames to acquire the crystal orientation matrix.
An equivalent polyhedron was then constructed to enable face normal vector determination.
The assigned face normal vectors were then used to align the crystallographic axes of our macroscopic crystal to the experimental reference frame for measurement.

\paragraph{Mechanical considerations --}
\label{sec:mech}
Because of the piezomagnetic effect, mechanical stress can affect magnetic and transport properties of MnTe crystals.
Therefore, in measuring these properties, the mounting of the sample has to minimize the stress while ensuring required position stability.
We thus used Apiezon M and H, and Toray Hivac silicone greases for mounting the samples.
We also note that the crystals rigidly fixed by G-varnish were often found fractured after thermal cycles in magnetic fields.
This behavior is likely the result of substantial stress originating from the combined effect of piezomagnetism and magnetostriction \cite{Note2}.

\paragraph{Transport, thermodynamic, and magnetic properties --}
\label{sec:transp-mang-prop}
The transport properties, heat capacity, and magnetization of MnTe crystals were measured at AU.
For transport measurements, the crystal were cut to a rectangular bar shape along the $x$-axis and mirror-polished.
The larger faces of the bar are parallel to the $(001)$ crystallographic face.
A crystal bar was used as the sample for both electric and thermoelectric transport measurements.
In resistivity and Hall measurements, the sample was loosely mounted to a sapphire substrate by high vacuum grease.
Standard six-point-probe measurements were carried out using a Quantum Design (QD) Electric Transport Option and Physical Properties Measurements System (PPMS).

For measuring Seebeck and Nernst effects, the sample was mounted as a heat-conducting bridge between two $\SI{1}{\kilo\ohm}$ resistors [Fig.~\ref{fig:complex}(b)].
Two $\phi\,\SI{25}{\micro \meter}$ E-type thermocouples made from Omega wires and two pairs of $\phi\,\SI{25}{\micro \meter}$ manganin wires were attached on the sample by conducting carbon paste to detect the temperature difference, and the Seebeck and the Nernst voltages, respectively.
The whole setup was fitted to a QD puck and inserted into the PPMS sample chamber.
In the measurement, the direction of the heat current was alternated, and the collected Seebeck and Nernst voltage were delta-averaged to eliminate the errors arising from the change of magnetic fields and the parasitic thermoelectric voltage.
Four Keithley 2182-A/J Nanovoltmeters and a Keithley 2600B source meter were used to detect the signals and to control the heat current, respectively.
The temperature and magnetic field was controlled via the commercial software of the PPMS.
All magnetotransport properties were measured in a transverse geometry, in which the magnetic field and the electric or heat current were applied along the $c$-axis and the $x$-axis, respectively.
The magnetic field scans followed the cycle $\SI{0}{\tesla}\rightarrow\SI{9}{\tesla}\rightarrow\SI{-9}{\tesla}\rightarrow\SI{0}{\tesla}$.

The heat capacity of the samples were measured with the heat capacity option of the PPMS.
Magnetic properties were investigated with the help of a QD MPMS-3 magnetometer.
After crystal orientation, a polished platelet crystal was attached to a quartz paddle by Apiezon grease with the magnetic field either along $[10\bar{1}0]$ or $c$-direction \cite{Note2}.

Although MnTe shows a polymorphic transition at $T > \SI{600}{\kelvin}$ \cite{mori2020}, no trace of structural distortion or twinning can be detected at $T\leq \SI{400}{\kelvin}$ from both our SCXRD and neutron studies \cite{baral2022}.
We thus assume that the observed phenomena are governed by the thermal evolutions of the magnetic domain state.

\paragraph{XAS and XMCD  -- }
\label{sec:xas-xmcd}

XAS and XMCD spectra was collected by total electron yield (TEY) mode at beamline BL-14U, NanoTerasu synchrotron facility, Sendai, Japan.
The polished sample was loosely mounted by conducting carbon tape on a copper sample holder such that the $c$-axis is parallel to the incident X-ray direction.
In XMCD measurements, the external magnetic field is applied parallel to the X-ray incident direction, which is parallel to the $c$-axis.
The XMCD signals are obtained by taking the difference between XAS measured using right- and left-circularly polarized X-rays.

For spectral calculations within the atomic model, 
the initial state was obtained by diagonalizing the multi-electron Hamiltonian 
constructed in the Fock basis for the $2p^6 3d^5$ configuration using the Lanczos method \cite{Sasabe2021, Sasabe2023}. 
Based on dipole transitions from this initial state, 
the XAS and XMCD spectra for the $2p^5 3d^6$ electronic configurations were evaluated.
The spin--orbit coupling constants and Slater integrals were obtained from ionic calculations 
based on the Hartree--Fock--Slater (HFS) method \cite{cowan1981}. 
For the Slater integrals, $\SI{70}{\percent}$ of the HFS values were used. 
The atomic model includes crystal-field parameters for $D_{3d}$ symmetry, 
with $B^{(2)}_{0} = -\SI{0.08}{\electronvolt}$, $B^{(4)}_{0} = \SI{0.83}{\electronvolt}$, and $B^{(4)}_{3} = \SI{0.10}{\electronvolt}$ \cite{Sasabe2021}.
In this study, we introduce effective internal fields of $\SI{0.30}{\electronvolt}$ arising from the ALTM interaction 
and $\SI{1}{\milli\electronvolt}$ from the weak ferromagnetic interaction.
\paragraph{NMR and EPR --}
\label{sec:nmr-epr}
The $^{55}$Mn NMR experiments were performed in a variable temperature insert of an Oxford Instruments $\SI{16}{\tesla}$ magnet using a home-built NMR probe.
The sample was mounted inside a tightly fitting three-turn NMR coil wound from thin copper wire embedded in a hardened epoxy/ZrO$_2$ composite.
The tuning and matching capacitors (1-20~pF) were mounted at the bottom of the probe, close to the NMR coil, with the matching capacitor connected in parallel.
This configuration, together with a small number of coil turns, enabled reliable tuning up to $490$~MHz.
A typical $\pi/2$ pulse length was $1~\mu$s.
NMR signals were recorded using the standard spin-echo $\pi/2$-$\tau_d$-$\pi$-$\tau_d$ pulse sequence, with a typical delay of $\tau_d = 10$~$\mu$s between the $\pi/2$ and $\pi$ pulses.
Spin-lattice relaxation times $T_1$ were measured using an inversion recovery pulse sequence, $\varphi_{\rm i}$-$\tau$-$\pi/2$-$\tau_d$-$\pi$-$\tau_d$, where the inversion pulse $\varphi_{\rm i}\approx\pi$ was followed by a variable delay $\tau$ before the read-out echo.
The $T_1$ datasets, obtained from integrated outer satelitte-transition intensities, were analyzed using the standard relaxation model for a quadrupolar nucleus with $I=5/2$ monitored on the $3/2\leftrightarrow 5/2$ transition;
\begin{align*}
  \label{eq:1}
  m(\tau)=  1-(1+s) & \left [\,0.0286\exp{(-\tau/T_1)} \right.\\
                    & \left. + 0.214\exp{(-3\tau/T_1)} \right.\\
                    & \left. + 0.4\exp{(-6\tau/T_1)} \right. \\
                    & \left. + 0.286\exp{(-10\tau/T_1)} \right. \\
                    & \left. +0.071\exp{(-15\tau/T_1)} \right]\,,
\end{align*}
where $T_1$ is the spin-lattice relaxation time and $s$ is the inversion factor.
Because of the large NMR linewidths, the data were typically fitted using a stretched-exponential form with a stretching exponent of around $0.5$.

Low-temperature continuous-wave X-band EPR measurements were performed on a Bruker E500 EPR spectrometer operating at 9.4 GHz and equipped with a 4122SHQE cylindrical Bruker resonator. The EPR spectra were measured at 1 mW microwave power, while the modulation field was set to 0.5 mT amplitude and 100 kHz modulation frequency.
For precise sample temperature control ($\pm \SI{0.05}{\kelvin}$ ), an Oxford Instruments ESR900 cryostat and ITC503 temperature controller were used.
High-temperature continuous-wave X-band EPR measurements were performed using a home-built spectrometer equipped with a Varian E-101 microwave bridge operating at 9.3 GHz, a Bruker ER 4105DR double cavity resonator and a Eurotherm temperature controller.
The EPR spectra were measured at 5 mW microwave power, while the modulation field was set to 0.5 mT amplitude and 80 kHz modulation frequency.
In both low and high-temperature measurements, the single crystal was mounted on a quartz crystal holder, enabling the rotation of the sample around an axis lying in the $ab$ plane.

\end{document}


\title{Intrinsic single crystals of MnTe altermagnet \\
  -- Supplementary Materials
}

\label{sec:authorlist}

\author{Kim-Khuong Huynh${}^\ddag$}
\email{hkkhuong@chem.au.dk}

\author{Michael Anthony Quintero${}^\ddag$}

\affiliation{Department of Chemistry, Aarhus University, Langelandsgade 140, 8000 Aarhus C, Denmark}

\thanks{These authors contributed equally to this work.}
\author{Martin Klanj\v{s}ek}
\affiliation{Jozef Stefan Institute, Jamova cesta 39, 1000 Ljubljana, Slovenia}

\author{Tilen Knafli\v{c}}
\affiliation{Jozef Stefan Institute, Jamova cesta 39, 1000 Ljubljana, Slovenia}
\affiliation{Silicon Austria Labs, Europastra\ss{}e 12, 9524 Villach, Austria}

\author{Yuta Ishii}
\author{Norimasa Sasabe}
\affiliation{Center for Basic Research on Materials, National Institute for Materials Science (NIMS), Tsukuba 305-0047, Japan}

\author{Nhu-Quynh T. Phan}
\affiliation{Department of Chemistry, Aarhus University, Langelandsgade 140, 8000 Aarhus C, Denmark}

\author{Frej S{\o}ren Rattenborg}
\affiliation{Department of Chemistry, Aarhus University, Langelandsgade 140, 8000 Aarhus C, Denmark}


\author{Yuichi Yamasaki}
\affiliation{Center for Basic Research on Materials, National Institute for Materials Science (NIMS), Tsukuba 305-0047, Japan}
\affiliation{International Center for Synchrotron Radiation Innovation Smart, Tohoku University, Sendai 980-8577, Japan}
\affiliation{Center for Emergent Matter Science (CEMS), RIKEN, Wako 351-0198, Japan}

\author{Denis Ar\v{c}on}
\affiliation{Jozef Stefan Institute, Jamova cesta 39, 1000 Ljubljana, Slovenia}
\affiliation{Faculty of Mathematics and Physics, University of Ljubljana, Jadranska cesta 19, 1000 Ljubljana, Slovenia}

\author{Bo Brummerstedt Iversen}
\email{bo@chem.au.dk}
\affiliation{Department of Chemistry, Aarhus University, Langelandsgade 140, 8000 Aarhus C, Denmark}

\date{\today}


\maketitle

\clearpage

\tableofcontents

\clearpage

\section{X-ray absorption spectroscopy}
\label{sec:x-ray-absorption}

\begin{figure}[H]
  \centering
  \includegraphics[width = .4\textwidth]{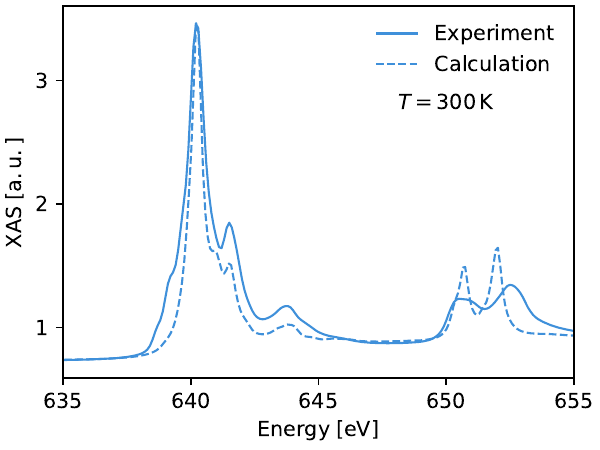}
  \caption{XAS spectrum of a MnTe collected at $T = \SI{300}{\kelvin}$ (solid line) in comparison to a theoretical calculation (dashed line).}
  \label{fig:xas}
\end{figure}

The X-ray absorption spectrum collected at $\SI{300}{\kelvin}$ and the theoretical calculation are shown in Fig.~\ref{fig:xas}.
The experimental measurement were carried out using total electron yield method.
The theoretical model takes into account the $2p^6 3d^5 \rightarrow 2p^5 3d^6$ dipole transition as described in the main text.

\section{Metastabilities around $\tn$ in resistivity and specific heat}
\label{sec:metastablities}
\begin{figure}[H]
  \centering
  \begin{subfigure}[t]{.45\textwidth}
    \includegraphics[width = \textwidth]{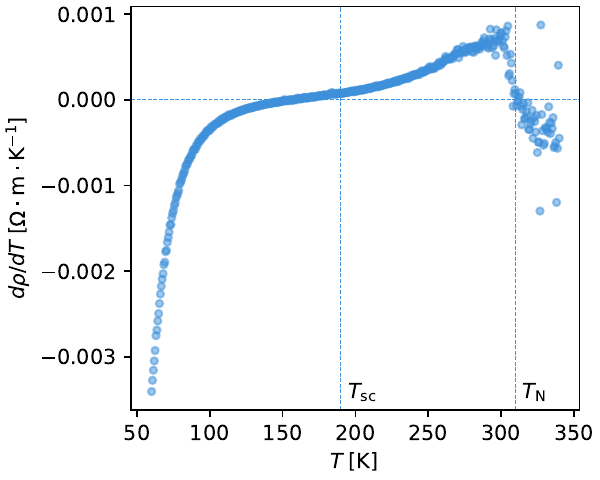}
    \subcaption{}
    \label{subfig:drhodT}
  \end{subfigure}
  \qquad
  \begin{subfigure}[t]{.45\textwidth}
    \includegraphics[width = \textwidth]{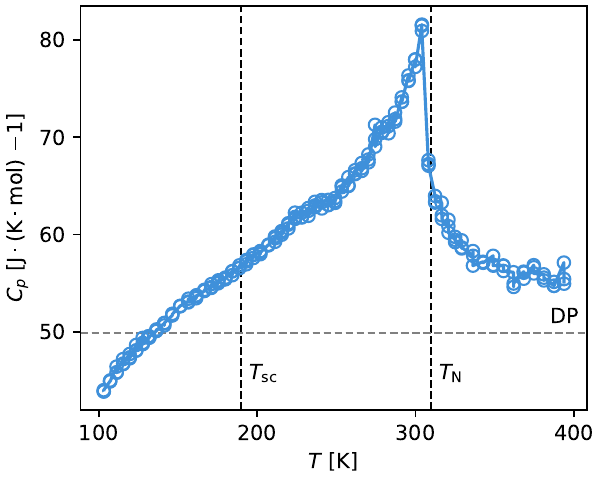}
    \subcaption{}
    \label{subfig:Cp}
  \end{subfigure}
  \caption{\subref{subfig:drhodT} Temperature-derivative of resistivity.
    \subref{subfig:Cp} $C_p$ around $\tn$.
  }
  \label{fig:drhodT-Cp}
\end{figure}

Additional signatures of the metastabilities emerging around $\tn$ can also be found in temperature dependence of the resistivity $\rho_{xx}$ and the specific heat $C_p$.
As shown in Fig.~\ref{fig:drhodT-Cp}, a wide shoulder appears in the supercooling regime between $\tn$ and $T_{\mathrm{sc}}$ of both the $T$-derivative of the resistivity $\rho_{xx}$ ($\partial \rho /\partial T$)and the specific heat $C_{\mathrm{p}}$ exhibits.
This feature centers at about $\SI{250}{\kelvin}$, yet extends over the whole supercooling window, being in a good agreement with the magnetic susceptibilities and the Sebeeck coefficient shown in the main text.

\section{Insulating regimes at $T <T_{\mathrm{min}}$}
\label{sec:low-temp-insul}

\begin{figure}[H]
  \centering
  \includegraphics[width = .45\textwidth]{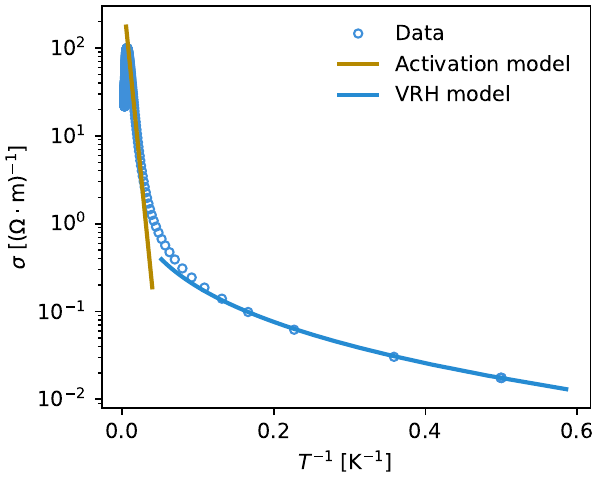}
  \caption{Temperature dependence of electrical conductivity (circles), showing activation and variable-range-hoping regimes (lines).}
  \label{fig:VRH}
\end{figure}

Fig.~\ref{fig:VRH} exhibits the insulating regimes at $T < T_{\mathrm{MI}} \approx \SI{150}{\kelvin}$ of MnTe.
In the temperature window $T_{\mathrm{MI}} < T < \SI{50}{\kelvin}$, the electrical conductivity $\sigma$ follows the thermal activation rule,
\begin{align}
  \label{eq:activation}
  \sigma(T) = \sigma_{0} \exp{(- E_{a}/k_{\mathrm{B}}T)}\,
\end{align}
where $E_a$ is the activation energy.
At lower temperatures, $T < \SI{20}{\kelvin}$, the conductivity crosses over to the variable-range-hopping regime;
\begin{align}
  \label{eq:vrh}
  \sigma(T) = \sigma_{0} \exp{[ - (T / T)^{1/4} ] }
\end{align}

Fitting $\sigma(T)$ data to these two models yields $E_a \approx \SI{17}{\milli \electronvolt}$ and $T_0 \approx \SI{5376}{\kelvin}$.

\section{Additional magnetization data}
\label{sec:addit-magn-data}

\begin{figure}
  \centering
  \begin{subfigure}[t]{.2\textwidth}
    \includegraphics[width = \textwidth]{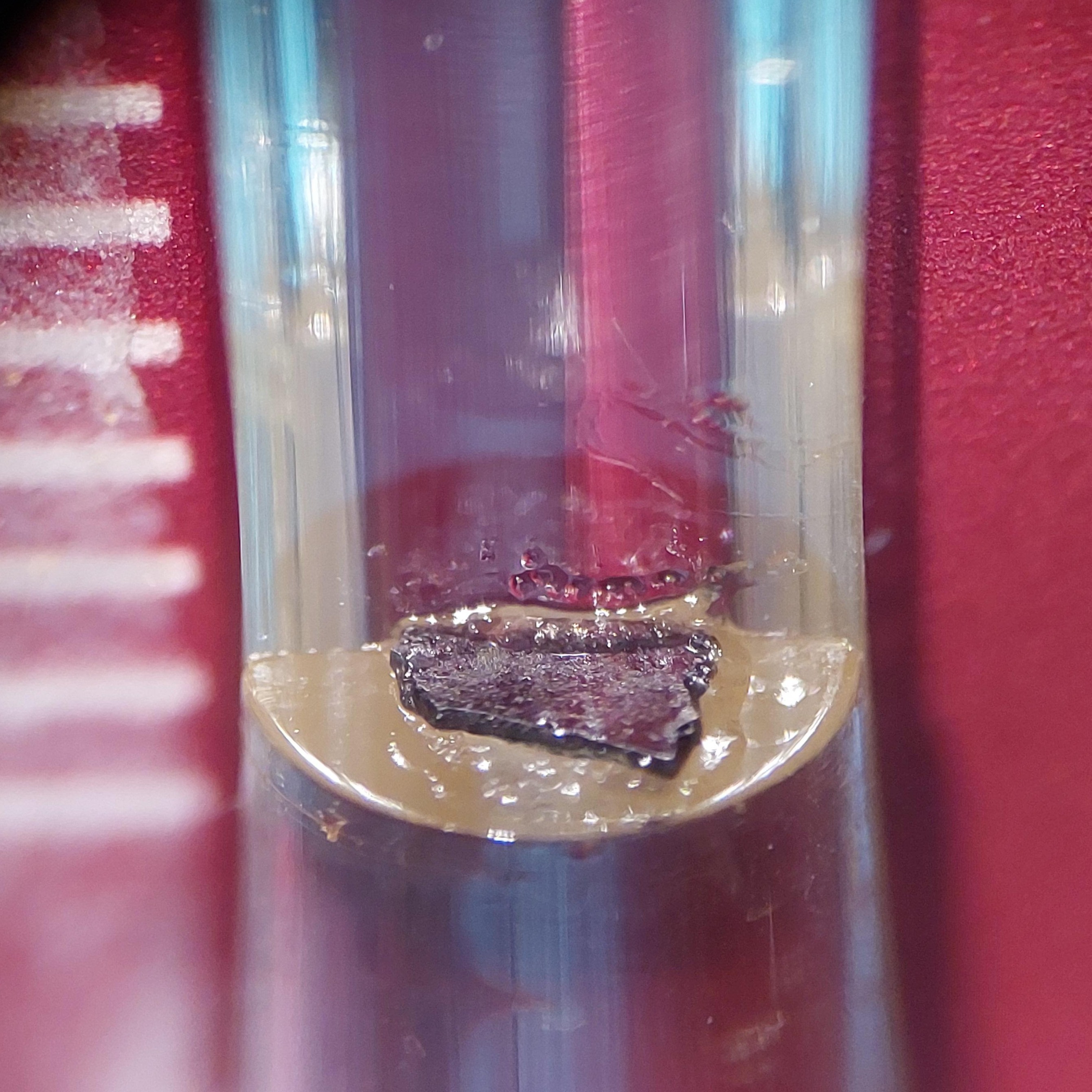}
    \subcaption{}
    \label{subfig:magrod-c}
  \end{subfigure}
  \qquad
  \begin{subfigure}[t]{.2\textwidth}
    \includegraphics[width = \textwidth]{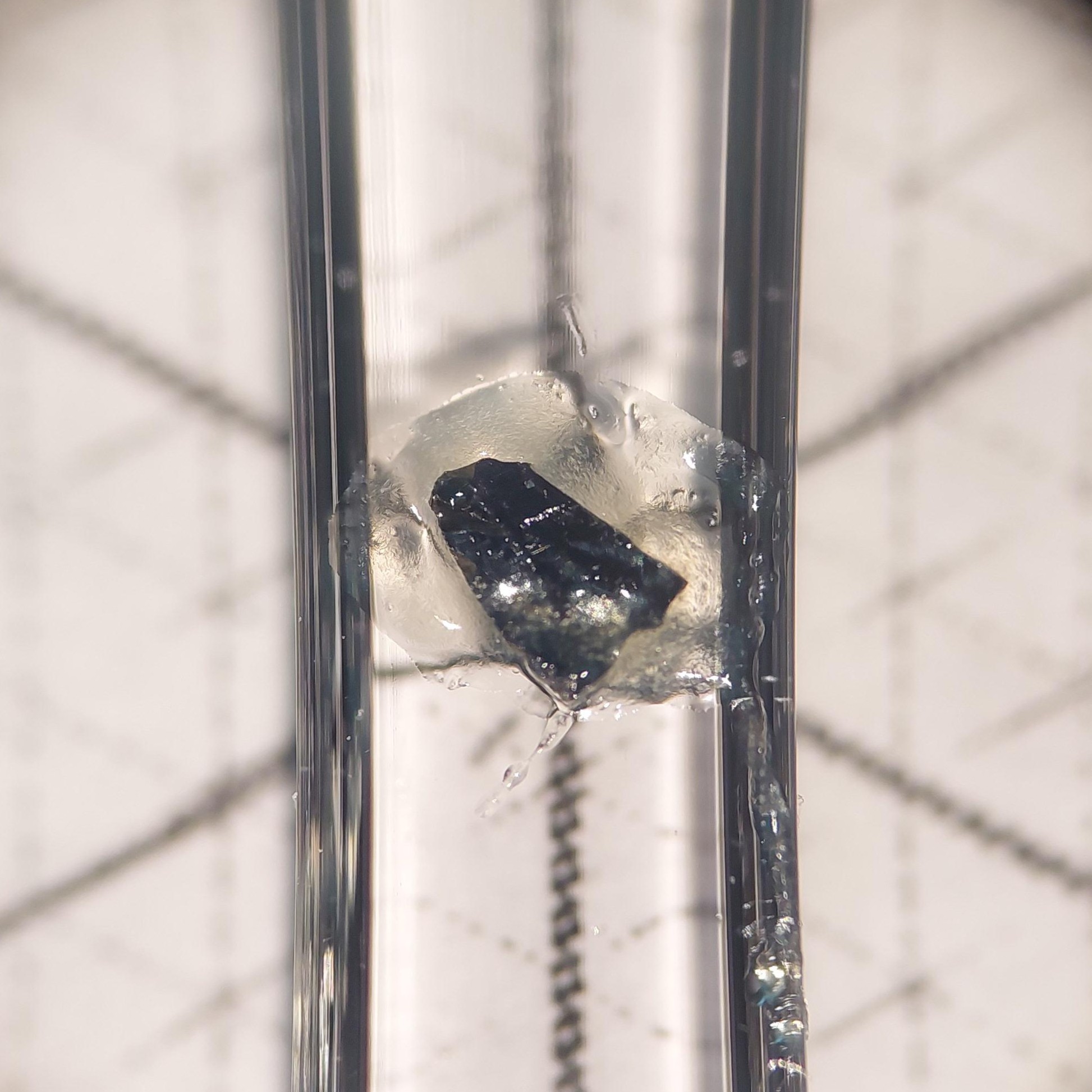}
    \subcaption{}
    \label{subfig:magrod-ab}
  \end{subfigure}
  \caption{
    A crystal of MnTe mounted on quartz rods for magnetization measurements with magnetic field parallel to \subref{subfig:magrod-c} $\hat{c}$-axis and \subref{subfig:magrod-ab} $[10\bar{1}0]$-direction.
    In each measurement, the magnetic field is parallel to the quartz paddle.
    The thickness and the width of the paddles are about $\SI{4}{\milli\meter}$ and $\SI{2}{\milli\meter}$, respectively.
  }
  \label{fig:magrods}
\end{figure}

The sample setups for the measurements of directional dependence of the magnetization is displayed in Fig.~\ref{fig:magrods}.
The employed sample holders are quartz paddles with low magnetic background provided by Quantum Design.
For the $\vec{H} \parallel \hat{c}$ measurement (Fig.~\ref{subfig:magrod-c}), the sample was placed on the flat end of a broken paddle, which was glued to the supporting rod by G-varnish.
For the $\vec{H} \parallel \hat{x}$ measurement (Fig.~\ref{subfig:magrod-ab}), the crystallographic orientation of the crystal was defined by single crystal diffraction, and it was mounted so that the applied magnetic field was parallel to $\hat{x}$- (or $[10\bar{1}0]$-) direction.
In both setups, a mixture of Apiezon M and H greases with an approximate volume ratio of $3:1$ was used to fix the sample to the holder.

\begin{figure}
  \centering
  \includegraphics[width = .4\textwidth]{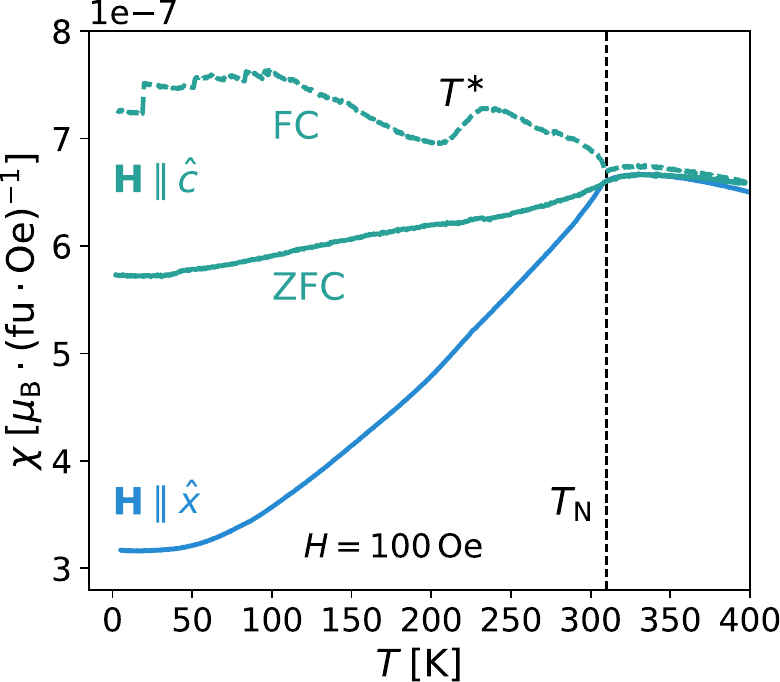}
  \caption{$\chi_{cc}$ and $\chi_{xx}$ measured in $H = \SI{100}{\oersted}$.}
  \label{fig:chi-100Oe}
\end{figure}

Fig.~\ref{fig:chi-100Oe} displays the magnetic susceptibility measured under $H = \SI{100}{\oersted}$ and in the whole temperature range $\SI{2}{\kelvin}$-$\SI{200}{\kelvin}$ allowed by the MPMS.
The susceptibilities show clear superheating and supercooling hysteresis.
The ZFCFW-FC bifurcation occurs at $T = \SI{400}{\kelvin}$.
Because this temperature is also the upper limit of all of our measurement systems, the superheating temperature $T_{\mathrm{sh}}$ remains hidden to our experiments.
On the other hand, the clear drop at $T_{\mathrm{sc}}\approx \SI{190}{\kelvin}$ in the FC branch of $\chi_{cc}$ is in a good agreement with the Seebeck data shown in the main text.

\begin{figure}
  \centering
  \begin{subfigure}[t]{.45\textwidth}
    \includegraphics[width = \textwidth]{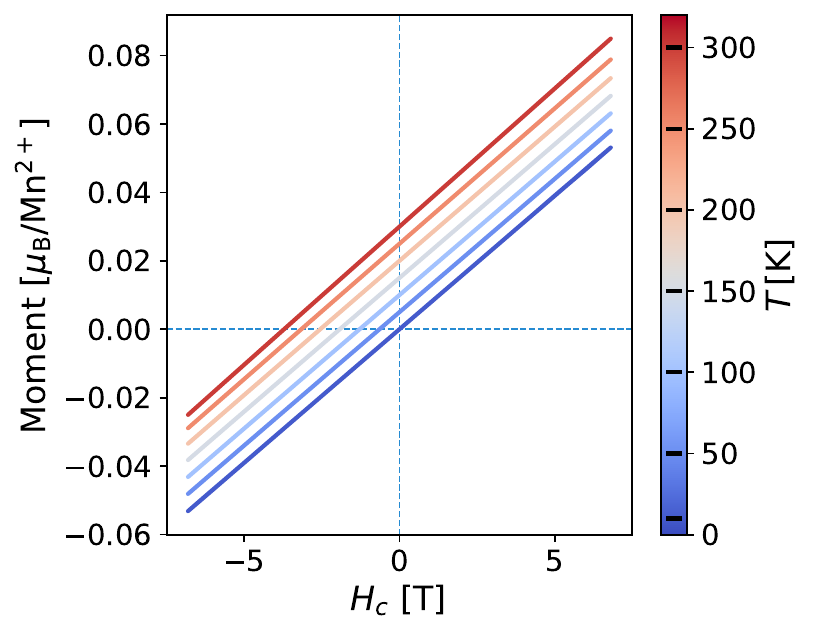}
    \subcaption{}
    \label{subfig:MHc}
  \end{subfigure}
  \qquad
  \begin{subfigure}[t]{.45\textwidth}
    \includegraphics[width = \textwidth]{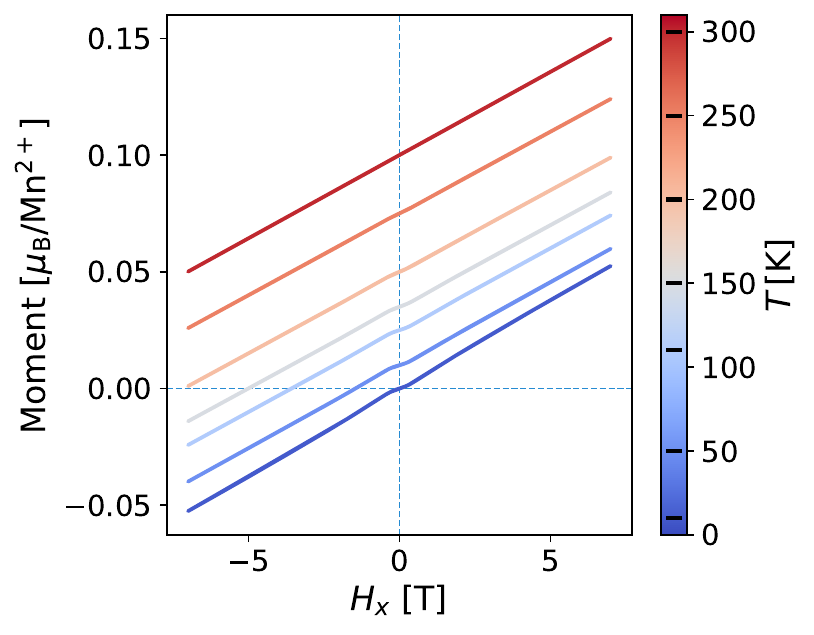}
    \subcaption{}
    \label{subfig:MHx}
  \end{subfigure}
  \caption{
    Magnetization versus magnetic field parallel to \subref{subfig:MHc} the $\hat{c}$-axis and \subref{subfig:MHx} $\hat{x}$-direction.
    For each curve, the magnetic field scans through the full cycle between $\SI{7}{\tesla}$ and $\SI{7}{\tesla}$, i.e., $\SI{0}{\tesla} \rightarrow \SI{7}{\tesla} \rightarrow \SI{-7}{\tesla} \rightarrow \SI{0}{\tesla} $.
    The curves are shifted for clarity.
  }
  \label{fig:MH}
\end{figure}

Fig.~\ref{subfig:MHc} and \ref{subfig:MHx} displays the magnetization curves measured at various temperatures in $\vec{H}_c$ and $\vec{H}_x$, respectively.
The $\hat{c}$-magnetization ($M_c$) curves are linear in the whole temperature range.
On the other hand, $M_x$ exhibits a mild slope bend at $|H_x| \approx \SI{0.5}{\tesla}$.
The non-linearity, which can be clearly seen at low temperatures, persists up to $T = \SI{300}{\kelvin}$, however with reduced magnitude.
In both orientations of the magnetic fields, the magnetization curves display negligible hysteresis with respect to the scanning magnetic fields.

\section{Magnetic resonance measurements}
\label{sec:nmr-measurements}

\subsection{Domain effects and the linewidth of NMR}
\label{sec:nmr}


The observed NMR linewidths are relatively broad, of the order of $2$~MHz, and we find that they show a noticeable dependence on the RF irradiation conditions.
Such a behaviour is naturally explained by the multidomain altermagnetic structure of MnTe. Different altermagnetic domains carry different orientations of the hyperfine field, leading to shifted resonance conditions for each domain.
In addition, the large internal hyperfine field defines the nuclear quantization axis, which is not perpendicular to the axis of the NMR coil used in the experiment.
As a result, the effective transverse RF field differs from one domain to another, so that different domains exhibit different optimal pulse lengths and excitation efficiencies.
The observed linewidth therefore reflects the superposition of signals from multiple domains with different RF coupling strengths, providing microscopic evidence for the multidomain altermagnetic structure inferred from
macroscopic measurements.
This geometry also explains why applying a small in-plane magnetic field significantly enhances the $^{55}$Mn NMR signal, as seen for the $10$~K spectrum in Fig. 4(a) of the main text, since it improves the RF excitation efficiency in the presence of the strong internal field.


\subsection{Possiblity of a spin-flop transition}
\label{sec:poss-spin-flop}
As the magnetization would completely reorient in the spin-flop phase, that should also dramatically change the local field sensed by the given ${}^{55}$Mn nucleus.
In general, the local field is given by the product $\vec{A}\cdot\vec{S}$, where $\vec{A}$ is the second rank hyperfine coupling tensor and $\vec{S}$ is the electronic spin.
Therefore, if the electronic spin vector $\vec{S}$ changes its direction, it couples to different components in the hyperfine coupling tensor and thus the effective field at the Mn nuclear site should also change \cite{slichter2013}.
As a result, the ${}^{55}$Mn NMR resonance line should be found at some other resonant frequency.
Such frequency shift are not observed in our measurement, and therefore the spin-flop transition is unlikely the mechanism for the change in the NMR spectral shape and the nonlinearity of the in-plane magnetization curves.

\subsection{EPR}
\label{sec:epr}

\begin{figure}
  \centering
  \includegraphics[width=.5\textwidth]{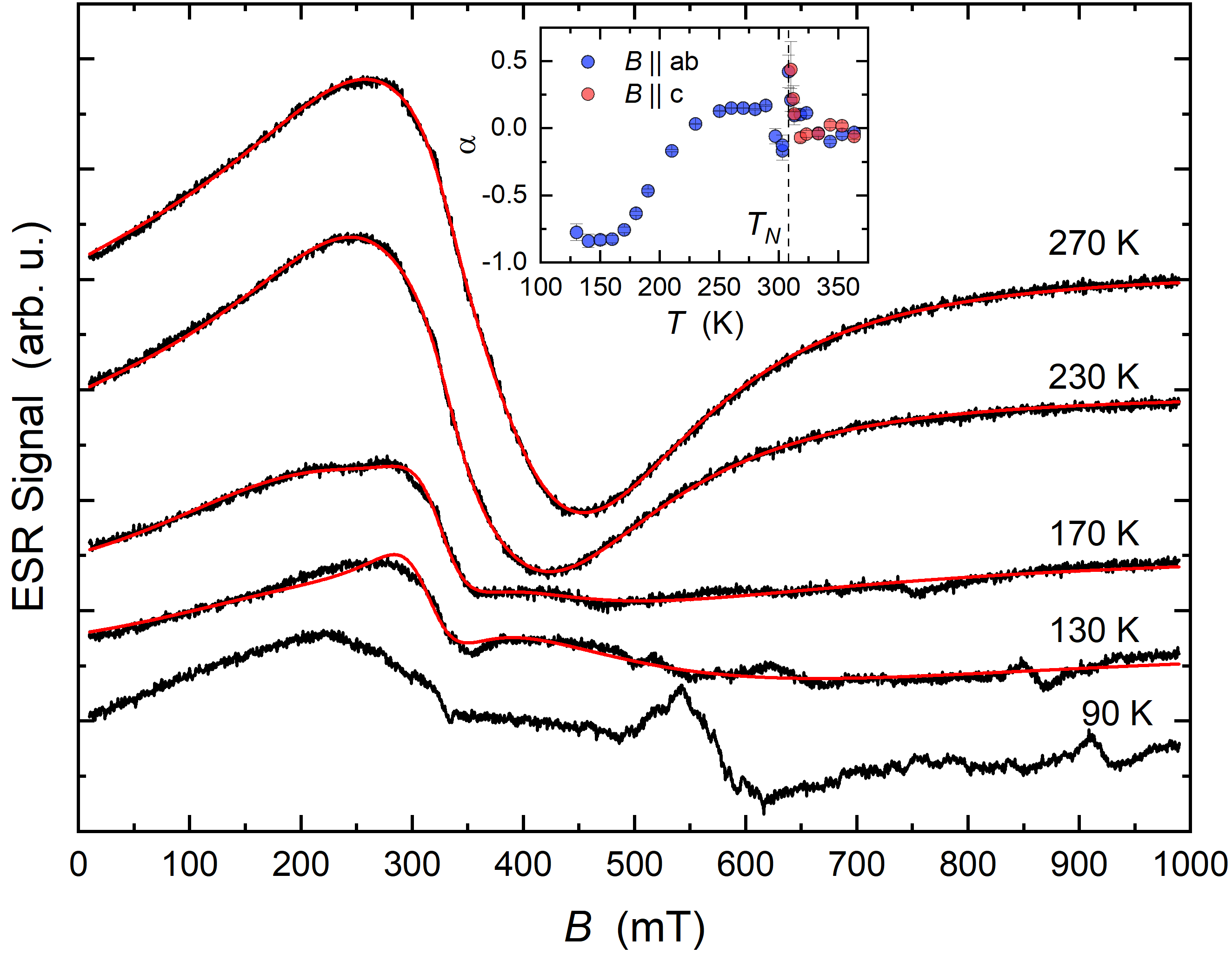}
  \caption{ EPR spectra at selected temperatures.The solid black curves represent the experimental data, while the solid red curves are fits consisting of two contributions: a broad Dysonian component associated with the intrinsic EPR response and a narrow extrinsic Lorentzian component arising from the sample holder.
    The inset shows the temperature dependence of the Dyson lineshape parameter $\alpha$ for the two perpendicular magnetic-field orientations.}
  \label{fig:EPR}
\end{figure}

The existence of Anderson-localized states is supported by X-band EPR measurements.
Well above the N\'eel temperature, the  EPR spectra at $g \approx 2$ are very broad and of slightly asymmetric lineshape, as shown in Fig.~\ref{fig:EPR}.
With temperature approaching $T_{\rm N}$ from above, the  EPR spectra  undergo a critical broadening with the  EPR linewidth diverging as $\Delta B\propto (T-T_{\rm N})^{-1}$, see Fig. 2(f) in the main text.
Below $T_{\rm N}$ in the out-of-plane field orientation, X-band EPR signal completely disappears as expected for the antiferromagnetically ordered system where antiferromagnetic resonance modes can be detected only at very high resonant frequencies \cite{povarov2025}.
However, for the in-plane field orientation, the residual high-temperature X-band EPR signal can still be traced well below $T_{\rm N}$.
For this configuration, the residual EPR signal linewidth initially decreases with decreasing temperature and reaches a minimum  value of around 330 mT at 240 K, which is a temperature that coincides with $T^{\ast}$, where a hysteresis loop in the Seebeck coefficient is observed [Fig. 2(b) of the main text].
On further cooling below $T^{\ast}$, the X-band EPR spectra start to broaden again and the EPR linewidth increases linearly with decreasing temperature. Moreover, the EPR spectra become more complex also with several additional components becoming increasingly more prominent with decreasing temperature [Fig.~\ref{fig:EPR}].
However, the extreme broadening of X-band EPR spectra below the metal-to-insulator temperature prevents us to accurately fit the EPR signal. 

The residual X-band EPR signal showing dramatic changes at the temperatures, where the charge transport also changes, suggests that the signal is due to the depleted states at the Fermi energy.
Such itinerant states give rise to the conducting electron spin resonance (CESR) characterized by the asymmetric Dysonian lineshape \cite{dyson1955}.
Therefore, to account for the observed EPR lineshape asymmetry at all temperatures, the spectra were fitted using the Dyson function, where the lineshape parameter $\alpha$ measures the signal asymmetry.
As shown in the inset of Fig.~\ref{fig:EPR}, in the temperature dependence of $\alpha (T)$ we observe that with decreasing temperature, the absolute value of $\alpha$ changes from the high temperature value of $0.14$ to $-0.84$ at $\SI{150}{\kelvin}$ corroborating the metallic state of MnTe in this temperature range.
rom the high temperature value of 0.14 to -0.84 at 150 K.
However, when the states contributing to the spectral weight at the Fermi energy lose their spatial coherence at the Anderson localization temperature, the CESR lineshape become anomalous as indeed experimentally detected below $T_{\rm MI}$. 

\section{Frangibility of  \element{Mn}\element{Te} single crystals}
\label{sec:mnte-fragibility}

\begin{figure}
  \centering
  \begin{subfigure}[t]{.2\textwidth}
    \includegraphics[width = \textwidth]{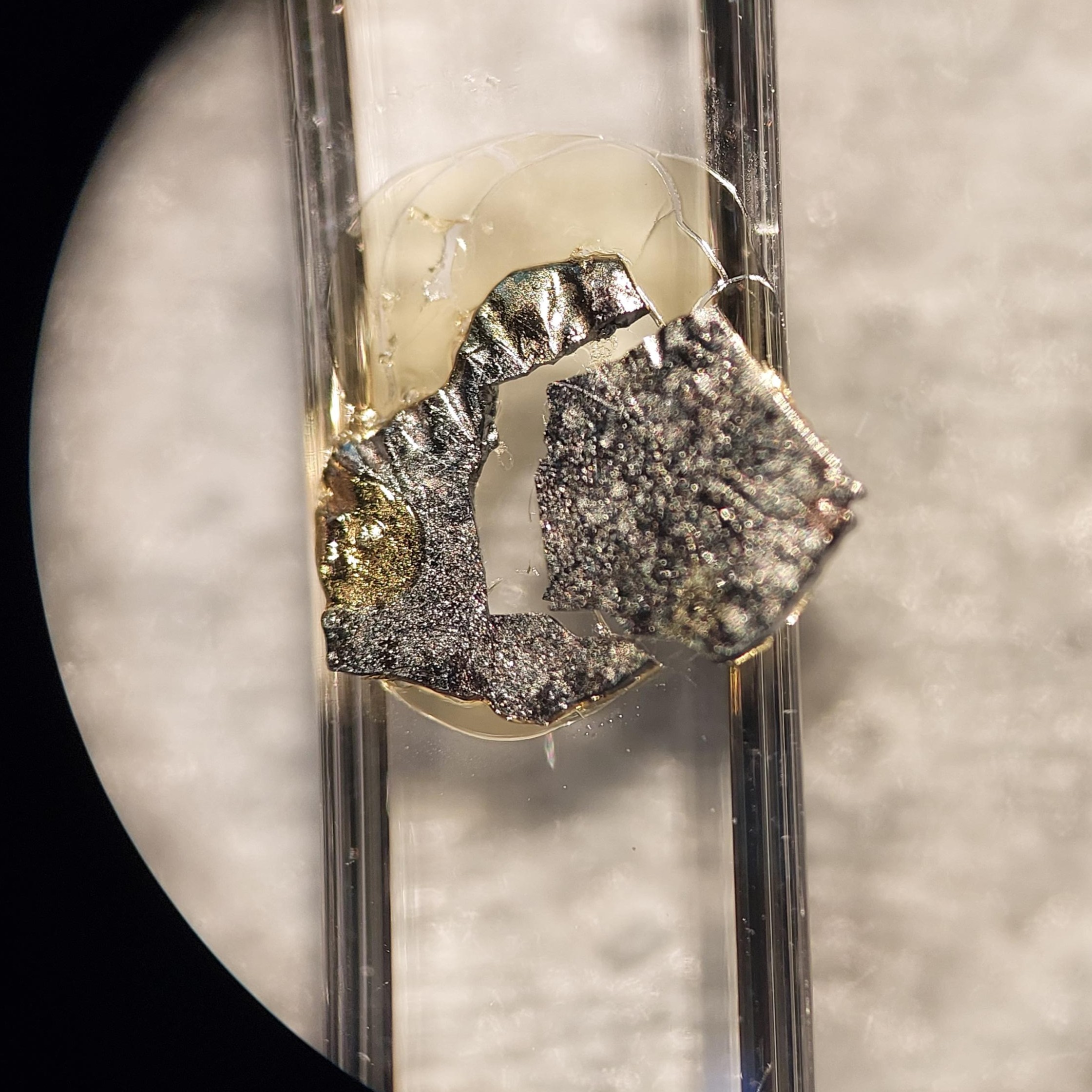}
    \subcaption{}
    \label{subfig:broken-mag}
  \end{subfigure}
  \qquad
  \begin{subfigure}[t]{.2\textwidth}
    \includegraphics[width = \textwidth]{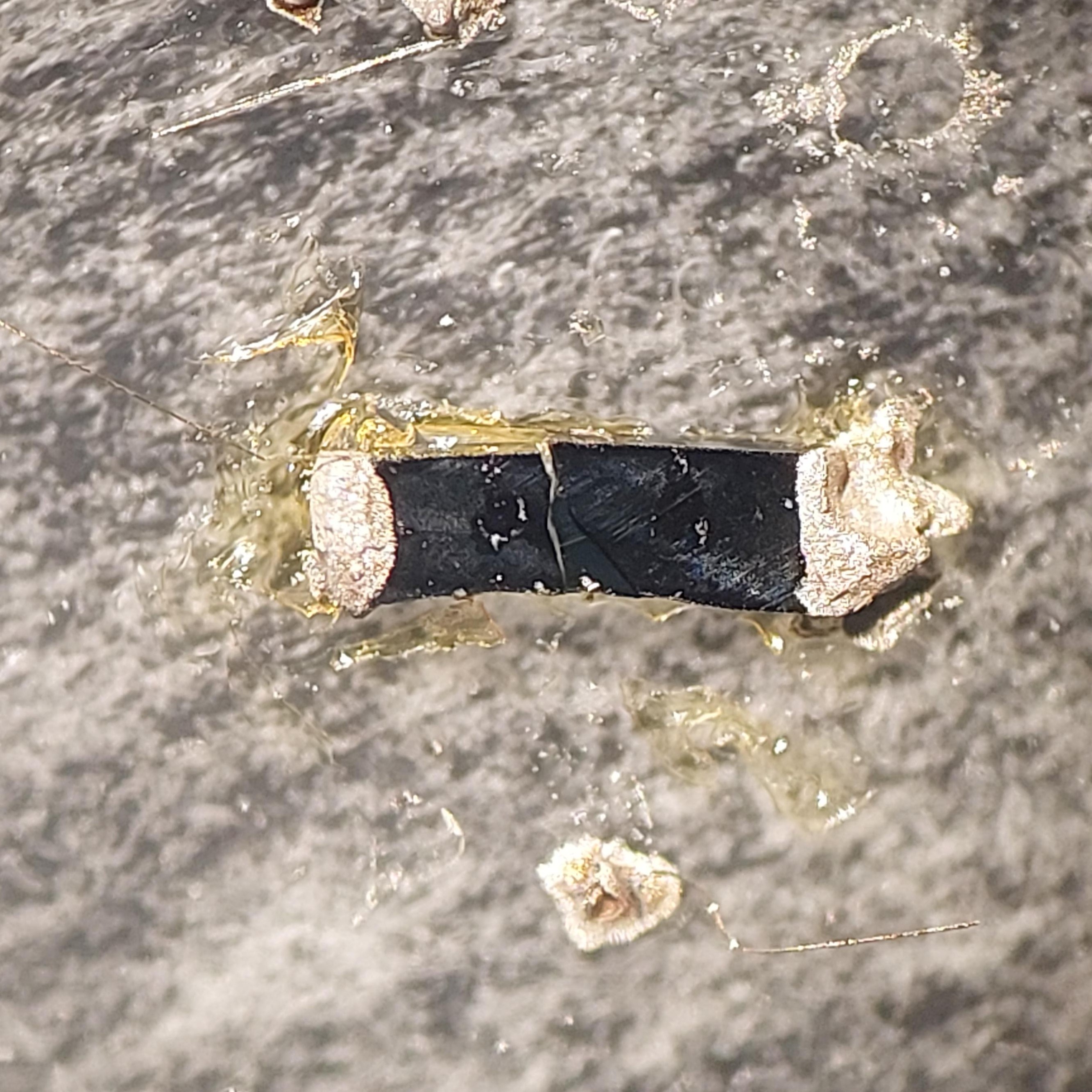}
    \subcaption{}
    \label{subfig:broken-transport}
  \end{subfigure}
  \caption{MnTe crystals fractured after the measurements of \subref{subfig:broken-mag} magnetic properties and \subref{subfig:broken-transport} transport properties.
   The crystals were attached by G-varnish to a quartz rod and a sapphire substrate in \subref{subfig:broken-mag} and \subref{subfig:broken-transport}, respectively.
  }
  \label{fig:broken-xtal}
\end{figure}

As shown in Fig.~\ref{fig:broken-xtal}, single crystals of MnTe were found fractured if they were rigidly adhered to a substrate during measurements with varying temperatures and magnetic fields.
This phenomenon indicates that MnTe is very fragile with respect to stress induced by mismatches in thermal expansion, and magnetostriction and piezomagnetic effects.
To avoid sample breaking, high-vacuum grease, e.g., Apiezon M, can be used to position the sample on the substrate, as displayed in Fig.~\ref{fig:magrods}.

\section{Cleaning the crystals from \element{Mn}\element{Te}$_2$ surface residues}
\label{sec:MnTe2-residues}
\begin{figure}
  \centering
  \begin{subfigure}[t]{.5\textwidth}
    \includegraphics[width = \textwidth]{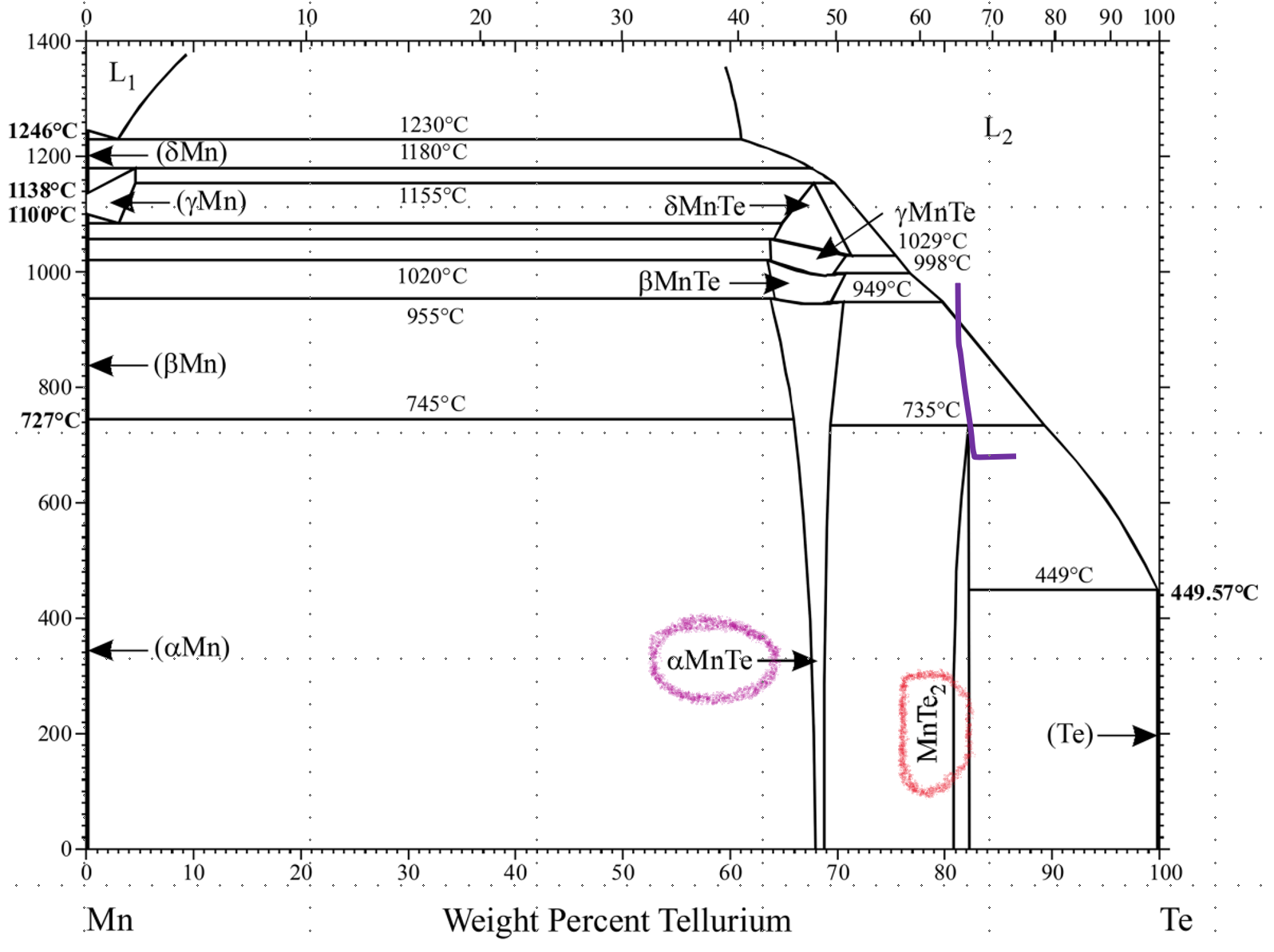}
    \subcaption{}
    \label{subfig:MnTe-phase-diagram}
  \end{subfigure}
  \qquad
  \begin{subfigure}[t]{.2\textwidth}
    \includegraphics[width = \textwidth]{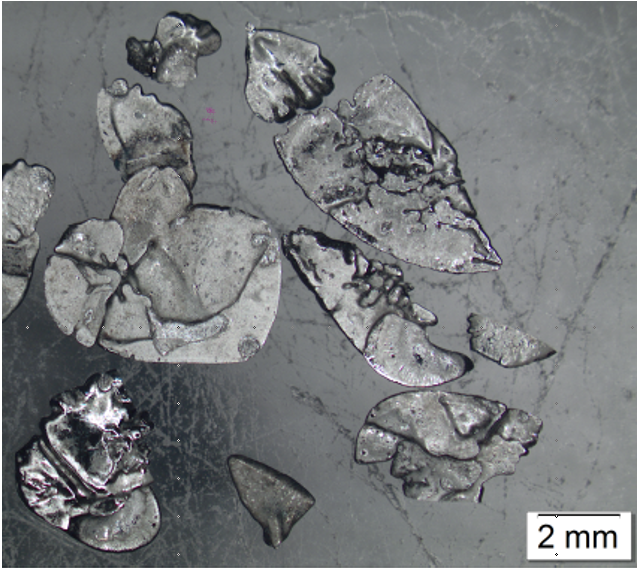}
    \subcaption{}
    \label{subfig:MnTe-as-grown}
  \end{subfigure}
  \caption{\subref{subfig:MnTe-phase-diagram} Mn-Te binary phase diagram obtained from Ref. \onlinecite{villars}.
    The purple line indicates the employed synthesis condition and heating program.
    The circles indicate the desired phase MnTe and the secondary phase MnTe$_2$.
    \subref{subfig:MnTe-as-grown} As grown MnTe single crystals.
    The dull color of the surface is due to the thin layer of Te-rich phases.
  }
  \label{fig:synthesis-as-grown}
\end{figure}

In the synthesizing MnTe crystals, Te is employed the self-flux to provide a Te-rich nucleation and growth condition essential in avoiding the formation of Te-vacancies.
However, the Te-rich flux can also be favorable for the formation of the secondary phase MnTe$_2$.
The formation of this phase can be mitigated by using the starting composition and decanting temperature shown in Fig.~\ref{subfig:MnTe-phase-diagram}.
As shown in Fig.~\ref{subfig:MnTe-as-grown}, as-grown MnTe crystals resulted from the selected synthesis conditions are thick platelets with about $\SI{5}{\milli\meter}$ in lateral dimensions.
The undesirable MnTe$_2$ phase only exist as the thin layer of gray color on the surfaces of the as-grown crystals.
No pockets of Te-rich phases were found in the bulk of the crystals by multiple cutting, continuous polishing, or extended acid-etching the crystals.

In order to remove the MnTe$_2$ residues on the surface, we soaked the as-grown crystals in diluted aqua regia for a few minutes.
After that, the crystals were washed a few times with deionized water to remove the excess acid and dried in air.
The washed crystals should be stored in vacuum or inert atmosphere, as MnTe surface degrades in the time scale of months in air.
The crystals can be polished before the measurements of physical properties.

\begin{figure}
  \centering
  \includegraphics[width = .45\textwidth]{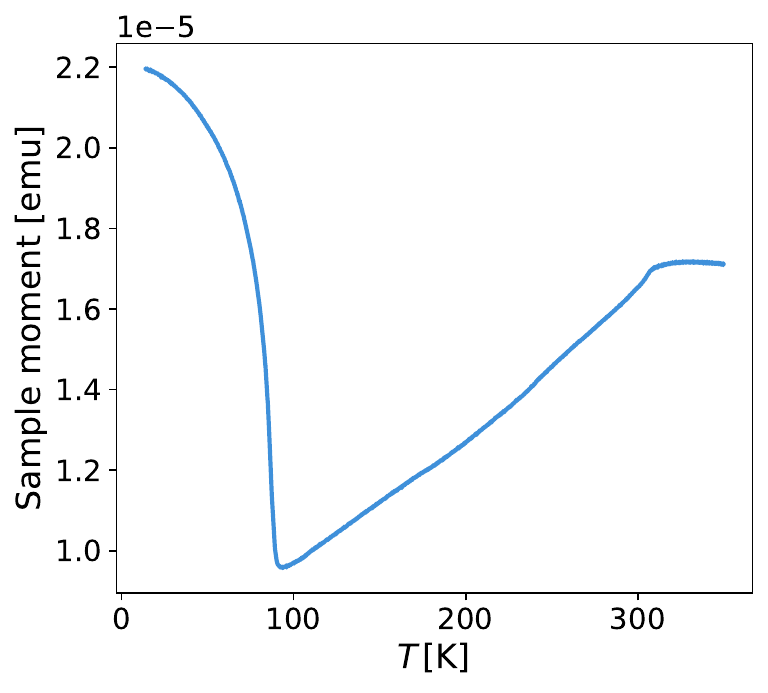}
  \caption{Total magnetic moment of an as-grown MnTe single crystal with MnTe$_2$ on the surface.
    The AFM transition of the MnTe$_2$ surface residues occurs at $T_{\mathrm{FM}} \approx \SI{90}{\kelvin}$.
    The magnetic field is $H = \SI{200}{\oersted}$ and is parallel to the $ab$ plane.}
  \label{fig:chi-contaminated}
\end{figure}

A MnTe crystal contaminated by MnTe$_2$ will exhibit a strong ferromagnetic-like transition at $T_{\mathrm{FM}} \approx \SI{90}{\kelvin}$ as displayed in Fig.~\ref{fig:chi-contaminated}.
This behavior, although appears ferromagnetic-like, originates from an unconventional antiferromagnetic ordering of MnTe$_2$ \cite{quintero2026,zhu2024d}.
The same measurements carried out on clean crystals do not show this transition, see Fig.~\ref{fig:chi-100Oe}.

\clearpage
\bibliographystyle{apsrev4-1}


\bibliography{./04a_MnTe.bib}